\documentclass[pra, 12pt, showpacs, showkeys, eqsecnum]{revtex4}
\usepackage{graphics}

\begin{document}
\title{Probability sum rules and consistent quantum histories}
\author{Thomas F. Jordan}
\email[email: ]{tjordan@d.umn.edu}
\affiliation{Physics Department, University of Minnesota, Duluth, Minnesota 55812 USA}
\author{Eric D. Chisolm}
\email[email: ]{echisolm@lanl.gov}
\affiliation{Theoretical Division, Los Alamos National Laboratory, Los Alamos, 
New Mexico 87545 USA} 

\begin{abstract}
An example shows that weak decoherence is more restrictive than the
minimal logical decoherence structure that allows probabilities to be
used consistently for quantum histories. The probabilities in the sum rules that define minimal decoherence are all calculated by using a projection operator to describe each possibility for the state at each time. Weak decoherence requires more sum rules. They bring in additional variables, that require different measurements and a different way to calculate probabilities, and raise questions of operational meaning. The example shows that extending the linearly positive probability formula from weak to minimal decoherence gives probabilities that are different from those calculated in the usual way using the Born and von Neumann rules and a projection operator at each time.
\end{abstract}


\maketitle

\section{Introduction}\label{one}

A basic property of consistent quantum histories, whether they are
used to construct an interpretation of quantum mechanics or to
understand emergence of classical behavior, is that there are
probability sum rules that describe decoherence; they give
probabilities as sums of probabilities without quantum interference
terms \cite{Griffiths1984,Omnes1988,Omnes1990,Omnes1992,DowkerHalliwell1992,
Griffiths1993,Gell-MannHartle1993,OmnesBookCh,Griffiths1996,GriffithsBookCh}. 
The minimal logical decoherence structure for consistent quantum
histories was described by Omn\`{e}s
\cite{Omnes1988,Omnes1990,OmnesBookCh}. We will call it
\textit{minimal decoherence}. A different structure, which is easier
to describe and use, is called \textit{weak decoherence}
\cite{Gell-MannHartle1993}. It appears to be not completely
appreciated that the two are different, that minimal decoherence does
not imply weak decoherence as an outline of a proof too readily
suggests \cite{DowkerHalliwell1992}, and that the sum rules that do
imply weak decoherence \cite{Gell-MannHartle1993} involve more
far-reaching assumptions.

Here we give a simple example that makes this clear. The probabilities
in the example satisfy the sum rules that define minimal decoherence
but not those needed for weak decoherence. In minimal decoherence,
each sum rule is for a probability, calculated using the Born and von
Neumann rules, for the truth at successive times of propositions
represented by projection operators. The sum rules of
minimal decoherence relate all the probabilities that can be
calculated this way, the probabilities for all possible outcomes of measurements that test those propositions, so minimal decoherence is consistent within
itself. Weak decoherence requires more sum rules. It allows fewer
histories. It brings in new variables that require new measurements
and a new way to calculate probabilities. It involves more assumptions
than are necessary. It raises questions of operational meaning. There
is an altogether larger project \cite{Isham94,IshamLinden94} we need
not pursue. We will discuss this in the final section.

In the proposal of linearly positive histories and probabilities
\cite{GoldsteinPage95,Hartle04}, the foundation of the formula for
probabilities is that it agrees with the probabilities calculated in
the usual way for histories that satisfy the sum rules for weak
decoherence \cite{GoldsteinPage95}. Our example shows that this
generally does not extend to histories that satisfy the sum rules for
minimal decoherence.  The linearly positive formula gives
probabilities for histories in the example that are different from the
probabilities calculated in the usual way using the Born and von
Neumann rules and a projection operator at each time. Our perspective
changes when we look beyond weak decoherence and bring minimal
decoherence into view.
 
\section{Histories and probabilities}\label{two}

We need to use only a simple set of quantum histories to make our
point. Consider a history where at successive times a quantum system
is in states represented by four different state vectors: first
$|0\rangle $, then $|j\rangle $, then $|k\rangle $, and finally
$|f\rangle $. Let $j\triangleright k\triangleright f$ denote this
history. We let $|j\rangle $ be one of two orthonormal vectors
$|1\rangle $ and $|2\rangle $, let $|k\rangle $ be one of three
orthonormal vectors $|3\rangle $, $|4\rangle $ and $|5\rangle $, and
let $|f\rangle $ be one of three orthonormal vectors $|6\rangle $,
$|7\rangle $ and $|8\rangle $. In the set we consider, there will be
no histories that go through $|3\rangle $ or $|4\rangle $ to
$|8\rangle $ or through $|5\rangle $ to $|6\rangle $ or $|7\rangle
$. Thus we consider ten $j\triangleright k\triangleright f$ histories:
\begin{center}
$1\triangleright 3\triangleright 6$, $\; \; 1\triangleright 4\triangleright 6$, $\; \; 2\triangleright 3\triangleright 6$, $\; \; 2\triangleright 4\triangleright 6$, \\
$1\triangleright 3\triangleright 7$, $\; \; 1\triangleright 4\triangleright 7$, $\; \; 2\triangleright 3\triangleright 7$, $\; \; 2\triangleright 4\triangleright 7$, \\
$1\triangleright 5\triangleright 8,\; \; $ and $\; \; 2\triangleright 5\triangleright 8$.
\end{center}

In the general formulation, there can be a different number of times,
the number of possibilities at each time can be larger or smaller, the
different possibilities at each time can be described by orthogonal
projection operators for subspaces of states instead of by particular
state vectors, the initial state can be mixed instead of pure, and all
possible histories can be considered. Nothing we need is omitted in
our simplification. The projection operators we use are simply formed
from the state vectors.

The probability for the history $j\triangleright k\triangleright f$ is
\begin{equation}
\label{probjk}
\langle f|k\rangle \langle k|j\rangle \langle j|0\rangle \langle 0|j\rangle \langle j|k\rangle \langle k|f\rangle .
\end{equation}
According to standard quantum mechanics, this is the probability, given the initial state represented by $|0\rangle $, for the truth at the successive times of the propositions represented by the projection operators $|j\rangle
\langle j|$, $|k\rangle \langle k|$ and $|f\rangle \langle f|$. These
projection operators are made to include the effect of the dynamics,
as in the Heisenberg picture.

Different histories $j\triangleright k\triangleright f$ are
independent, because at each time the different states are
independent. If the system goes through one of these histories, there
is no probability that it goes through another. Therefore, the
probability that the result is one or the other of two of these
histories, or of several, is the sum of the probabilities for those
histories. When we say here that the system goes through either the
history $j\triangleright 3\triangleright f$ or the history
$j\triangleright 4\triangleright f$, for example, we mean to imply
that it goes through either the state represented by $|3\rangle $ or
the state represented by $|4\rangle $, not a state represented by a
linear combination of $|3\rangle $ and $|4\rangle $.

When $|f\rangle $ is $|6\rangle $ or $|7\rangle $, we also consider
\begin{equation}
\label{probj34}
\langle f|\bigg(|3\rangle \langle 3|+|4\rangle \langle 4|\bigg)|j\rangle \langle j|0\rangle \langle 0|j\rangle \langle j|\bigg(|3\rangle \langle 3|+|4\rangle \langle 4|\bigg)|f\rangle .
\end{equation}
This is the probability that the system goes from $|0\rangle $ to
$|j\rangle $, and then to either $|3\rangle $ or $|4\rangle $ or a
state represented by a vector in the subspace spanned by $|3\rangle $
and $|4\rangle $, and finally to $|f\rangle $. It is the probability, given the initial state represented by $|0\rangle $, for the truth at the successive times of the
propositions represented by the projection operators $|j\rangle
\langle j|$, $|3\rangle \langle 3|+|4\rangle \langle 4|$ and
$|f\rangle \langle f|$. This can be when the system does not go
through either the history $j\triangleright 3\triangleright f$ or
$j\triangleright 4\triangleright f$. If we require that this
probability (\ref{probj34}) is just the sum of the probabilities for
the histories $j\triangleright 3\triangleright f$ and $j\triangleright
4\triangleright f$, we conclude that
\begin{equation}
\label{Rej34}
\textrm{Re}\langle f|3\rangle \langle 3|j\rangle \langle j|0\rangle 
\langle 0|j\rangle \langle j|4\rangle \langle 4|f\rangle  = 0
\end{equation}
for $j = 1,2$ and $f = 6,7$. For any $|f\rangle $, we consider the probability
\begin{equation}
\label{prob12k}
\langle f|k\rangle \langle k|\bigg(|1\rangle \langle 1| + |2\rangle \langle 2|\bigg)|0\rangle 
\langle 0|\bigg(|1\rangle \langle 1| + |2\rangle \langle 2|\bigg)|k\rangle \langle k|f\rangle ,
\end{equation}
given the initial state represented by $|0\rangle $, for the truth at the successive times of the propositions represented by the
projection operators $|1\rangle \langle 1|+|2\rangle \langle 2|$,
$|k\rangle \langle k|$ and $|f\rangle \langle f|$. If we require that
this is the sum of the probabilities for the histories
$1\triangleright k\triangleright f$ and $2\triangleright
k\triangleright f$, we conclude that
\begin{equation}
\label{Re12k}
\textrm{Re}\langle f|k\rangle \langle k|1\rangle \langle 1|0\rangle 
\langle 0|2\rangle \langle 2|k\rangle \langle k|f\rangle  = 0
\end{equation}
for $k = 3,4,5$ and $f = 6,7,8$. When $|f\rangle $ is $|6\rangle $ or
$|7\rangle $, we also consider the probability
\begin{equation}
\label{prob1234}
\langle f|\bigg(|3\rangle \langle 3|+|4\rangle \langle 4|\bigg)
\bigg(|1\rangle \langle 1| + |2\rangle \langle 2|\bigg)|0\rangle 
\langle 0|\bigg(|1\rangle \langle 1| + |2\rangle \langle 2|\bigg)
\bigg(|3\rangle \langle 3|+|4\rangle \langle 4|\bigg)|f\rangle ,
\end{equation}
given the initial state represented by $|0\rangle $, for the truth at the successive times of the propositions represented by the
projection operators $|1\rangle \langle 1|+|2\rangle \langle 2|$,
$|3\rangle \langle 3|+|4\rangle \langle 4|$ and $|f\rangle \langle
f|$. If we require that this is the sum of the probabilities
for the four histories $1\triangleright 3\triangleright f$,
$1\triangleright 4\triangleright f$, $2\triangleright 3\triangleright
f$, and $2\triangleright 4\triangleright f$, and assume that equations (\ref{Rej34}) and (\ref{Re12k}) hold, we conclude that
\begin{equation}
\label{Re1234}
\textrm{Re}\langle f|3\rangle \langle 3|1\rangle \langle 1|0\rangle 
\langle 0|2\rangle \langle 2|4\rangle \langle 4|f\rangle + 
\textrm{Re}\langle f|4\rangle \langle 4|1\rangle \langle 1|0\rangle 
\langle 0|2\rangle \langle 2|3\rangle \langle 3|f\rangle = 0
\end{equation}
for $f = 6,7$. These sum rules generally do not hold. The
probabilities (\ref{probj34}), (\ref{prob12k}) and (\ref{prob1234})
generally are not the same as the sums of the probabilities for the
individual histories, because the quantum interference terms
(\ref{Rej34}), (\ref{Re12k}), and (\ref{Re1234}) generally are not
zero. The condition that the interference terms are zero, so that the
sum rules do hold, is called \textit{decoherence}. These sum rules
relate probabilities calculated by using a projection operator to
describe each possibility for the state at each time. Sums of
probabilities correspond to sums of projection operators. We will call
these \textit{projection sum rules}. They ensure consistency among
these probabilities. They define the minimal logical structure of
decoherence with this consistency. We will call it \textit{minimal
decoherence}. It was described by Omn\`{e}s
\cite{Omnes1988,Omnes1990,OmnesBookCh}.

Weak decoherence \cite{Gell-MannHartle1993} is the condition that
\begin{equation}
\label{weak}
\textrm{Re}\langle f|k\rangle \langle k|j\rangle \langle j|0\rangle \langle 0|j'\rangle \langle j'|k'\rangle \langle k'|f\rangle  = 0 \; \, \textrm{unless} \; j=j' \; \textrm{and} \; k=k'.
\end{equation}
For the histories we are considering, this means that Eqs.\
(\ref{Rej34}), (\ref{Re12k}), and (\ref{Re1234}) hold and that each of
the two terms of Eqs.\ (\ref{Re1234}) is separately zero. Our example
will show that the latter is not implied by projection sum
rules. Minimal decoherence does not imply weak decoherence. A proof
that the two terms of Eqs.\ (\ref{Re1234}) are separately zero can be
made \cite{Gell-MannHartle1993} with an additional assumption: either
that
\begin{equation}
\label{prob?}
\langle f|\bigg(|3\rangle \langle 3|1\rangle \langle 1|+|4\rangle 
\langle 4|2\rangle \langle 2|\bigg)|0\rangle \langle 0|
\bigg(|1\rangle \langle 1|3\rangle \langle 3|+|2\rangle \langle 2|4\rangle \langle 4|\bigg)
|f\rangle 
\end{equation}
is the sum of the probabilities for the histories $1\triangleright
3\triangleright f$ and $2\triangleright 4\triangleright f$, or that
\begin{equation}
\label{prob??}
\langle f|\bigg(|4\rangle \langle 4|1\rangle \langle 1|+|3\rangle 
\langle 3|2\rangle \langle 2|\bigg)|0\rangle \langle 0|
\bigg(|1\rangle \langle 1|4\rangle \langle 4|+|2\rangle \langle 2|3\rangle \langle 3|\bigg)
|f\rangle 
\end{equation}
is the sum of the probabilities for the histories $1\triangleright
4\triangleright f$ and $2\triangleright 3\triangleright f$. These are
not projection sum rules. They are different. We will call them
\textit{history sum rules}. They include quantities, like
(\ref{prob?}) and (\ref{prob??}), that are not calculated by using
projection operators to describe the possibilities for the state at
each time. Our example will show that history sum rules are not
implied by projection sum rules. They are an additional assumption.

These additional sum rules distinguish weak from minimal
decoherence. They introduce a new kind of history, which cannot be
described with a
proposition represented by a projection operator at each time. Mathematically, they
are an inviting generalization, using operators to represent histories
and sums of operators for compositions of histories
\cite{Gell-MannHartle1993}, but the corresponding operational
procedures are not at all clear. We will consider this further in the
final section, but first we describe our example.

\section{Example}\label{three}

Our example is specified by the overlaps of its state vectors. Let
\begin{eqnarray}
\label{vectors}
|0\rangle  & = &  \frac{1}{\sqrt{2}}|1\rangle  +  \frac{i}{\sqrt{2}}|2\rangle  
\nonumber \\ 
|1\rangle  & = &  \frac{1}{2}|3\rangle  +  \frac{1}{2}|4\rangle + 
\frac{1}{\sqrt{2}}|5\rangle  \nonumber \\ 
|2\rangle  & = &  \frac{1}{2}|3\rangle  +  \frac{1}{2}|4\rangle - 
\frac{1}{\sqrt{2}}|5\rangle \nonumber \\ 
|6\rangle  & = &  \frac{1-i}{2}|3\rangle  +  \frac{1+i}{2}|4\rangle  \nonumber \\ 
|7\rangle  & = &  \frac{1+i}{2}|3\rangle  +  \frac{1-i}{2}|4\rangle  \nonumber \\ 
|8\rangle  & = &  |5\rangle.   
\end{eqnarray}
It is easy to see that Eqs.\ (\ref{Rej34}), (\ref{Re12k}), and
(\ref{Re1234}) are satisfied. The two terms of Eqs.\ (\ref{Re1234})
are not separately zero. They are
\begin{eqnarray}
\label{1234}
\langle 6|3\rangle \langle 3|1\rangle \langle 1|0\rangle \langle 0|2\rangle \langle 2|4\rangle \langle 4|6\rangle  & = & 1/16 \nonumber \\
\langle 6|4\rangle \langle 4|1\rangle \langle 1|0\rangle \langle 0|2\rangle \langle 2|3\rangle \langle 3|6\rangle  & = & -1/16 \nonumber \\
\langle 7|3\rangle \langle 3|1\rangle \langle 1|0\rangle \langle 0|2\rangle \langle 2|4\rangle \langle 4|7\rangle  & = & -1/16 \nonumber \\
\langle 7|4\rangle \langle 4|1\rangle \langle 1|0\rangle \langle 0|2\rangle \langle 2|3\rangle \langle 3|7\rangle  & = & 1/16 .
\end{eqnarray}
The projection sum rules are satisfied. The history sum rules are not.

The probability (\ref{probjk}) is $1/16$ for each of the four
histories $j\triangleright k\triangleright 6$ and $1/16$ for each of
the four histories $j\triangleright k\triangleright 7$; it is $1/4$
for each of the two histories $j\triangleright k\triangleright 8$. The
probability that a history goes to a particular $|f\rangle $ through
$|1\rangle $ is the same as the probability that it goes there through
$|2\rangle $, and the probability that a history goes to a particular
$|f\rangle $ through $|3\rangle $ is the same as the probability that
it goes there through $|4\rangle $. The probability that a history
goes to $|6\rangle $ is $1/4$, the probability that a history goes to
$|7\rangle $ is $1/4$, and the probability that a history goes to
$|8\rangle $ is $1/2$. From Eqs.\ (\ref{vectors}) we find that
\begin{equation}
\label{0from345}
|0\rangle  =  \frac{1+i}{2\sqrt{2}}|3\rangle  +  \frac{1+i}{2\sqrt{2}}|4\rangle +  \frac{1-i}{2}|5\rangle  
\end{equation}
and can see that
\begin{eqnarray}
\label{direct}
\langle 6|0\rangle \langle 0|6\rangle  & = & 1/4 \nonumber \\
\langle 7|0\rangle \langle 0|7\rangle  & = & 1/4 \nonumber \\
\langle 8|0\rangle \langle 0|8\rangle  & = & 1/2 .
\end{eqnarray}

The probabilities add up. Nothing has been left out. The vector
$|0\rangle $ is in the subspace spanned by $|1\rangle $ and $|2\rangle
$. A vector that would make a complete set of three orthonormal
vectors with $|1\rangle $ and $|2\rangle $ would be orthogonal to
$|0\rangle $, so it could not change anything we have considered.

In terms of probabilities, we can see directly that the history sum
rules are not satisfied. From Eqs.\ (\ref{1234}) and the probability
$1/16$ for each of the histories $j\triangleright k\triangleright 6$
and $j\triangleright k\triangleright 7$, we see that the quantity
(\ref{prob?}) is $1/4$ when $|f\rangle $ is $|6\rangle $ and zero when
$|f\rangle $ is $|7\rangle $, and the quantity (\ref{prob??}) is zero
when $|f\rangle $ is $|6\rangle $ and $1/4$ when $|f\rangle $ is
$|7\rangle $. The quantity (\ref{prob?}) is not the sum of the
probabilities for the histories $1\triangleright 3\triangleright f$
and $2\triangleright 4\triangleright f$, and the quantity
(\ref{prob??}) is not the sum of the probabilities for the histories
$1\triangleright 4\triangleright f$ and $2\triangleright
3\triangleright f$. These sums are both $1/8$.

We can also see directly that the projection sum rules do not imply
the history sum rules. The values of the quantities (\ref{prob?}) and
(\ref{prob??}) are changed when $|6\rangle $ and $|7\rangle $ are
interchanged. To check that the projection sum rules are satisfied, we
looked at the values of all the quantities in them. We can see that
none of these values are changed when $|6\rangle $ and $|7\rangle $
are interchanged. This confirms that the quantities (\ref{prob?}) and
(\ref{prob??}) in the history sum rules are not determined by
measurements of quantities in the projection sum rules.

\section{Linearly positive histories and probabilities}\label{four}

To consider the proposal of linearly positive histories and
probabilities \cite{GoldsteinPage95,Hartle04}, we look at
\begin{equation}
\label{lpprobjk}
p(j\triangleright k\triangleright f) = \textrm{Re}\langle 0|f\rangle \langle f|k\rangle \langle k|j\rangle \langle j|0\rangle .
\end{equation}
The proposal assumes that if $p(j\triangleright k\triangleright f)$ is
not negative, it is the probability for the history $j\triangleright
k\triangleright f$. For a set of histories that satisfies the
conditions (\ref{weak}) for weak decoherence, $p(j\triangleright
k\triangleright f)$ is nonnegative for every history $j\triangleright
k\triangleright f$ in the set and is the same as the probability
(\ref{probjk}) calculated in the usual way
\cite{GoldsteinPage95}. This is generally not true for a set of
histories that satisfies the conditions for minimal decoherence. For
the histories in our example, we find that $p(j\triangleright
k\triangleright f)$ is $1/8$ for each of the four histories
$1\triangleright 3\triangleright 6$, $2\triangleright 4\triangleright
6$, $2\triangleright 3\triangleright 7$, $1\triangleright
4\triangleright 7$ and is zero for each of the four histories
$2\triangleright 3\triangleright 6$, $1\triangleright 4\triangleright
6$, $1\triangleright 3\triangleright 7$, $2\triangleright
4\triangleright 7$. The probability (\ref{probjk}) is $1/16$ for each
of these eight histories. For each of the two histories
$1\triangleright 5\triangleright 8$ and $2\triangleright
5\triangleright 8$, we find that $p(j\triangleright k\triangleright
f)$ is $1/4$, which is the same as the probability (\ref{probjk}). The
proposal provides a simple formula for probabilities for weakly
decoherent histories, but it is not a reliable extension to other
histories.

\section{Trajectory graphs}\label{five}

\begin{figure}
\begin{center}
\scalebox{0.2}{\includegraphics{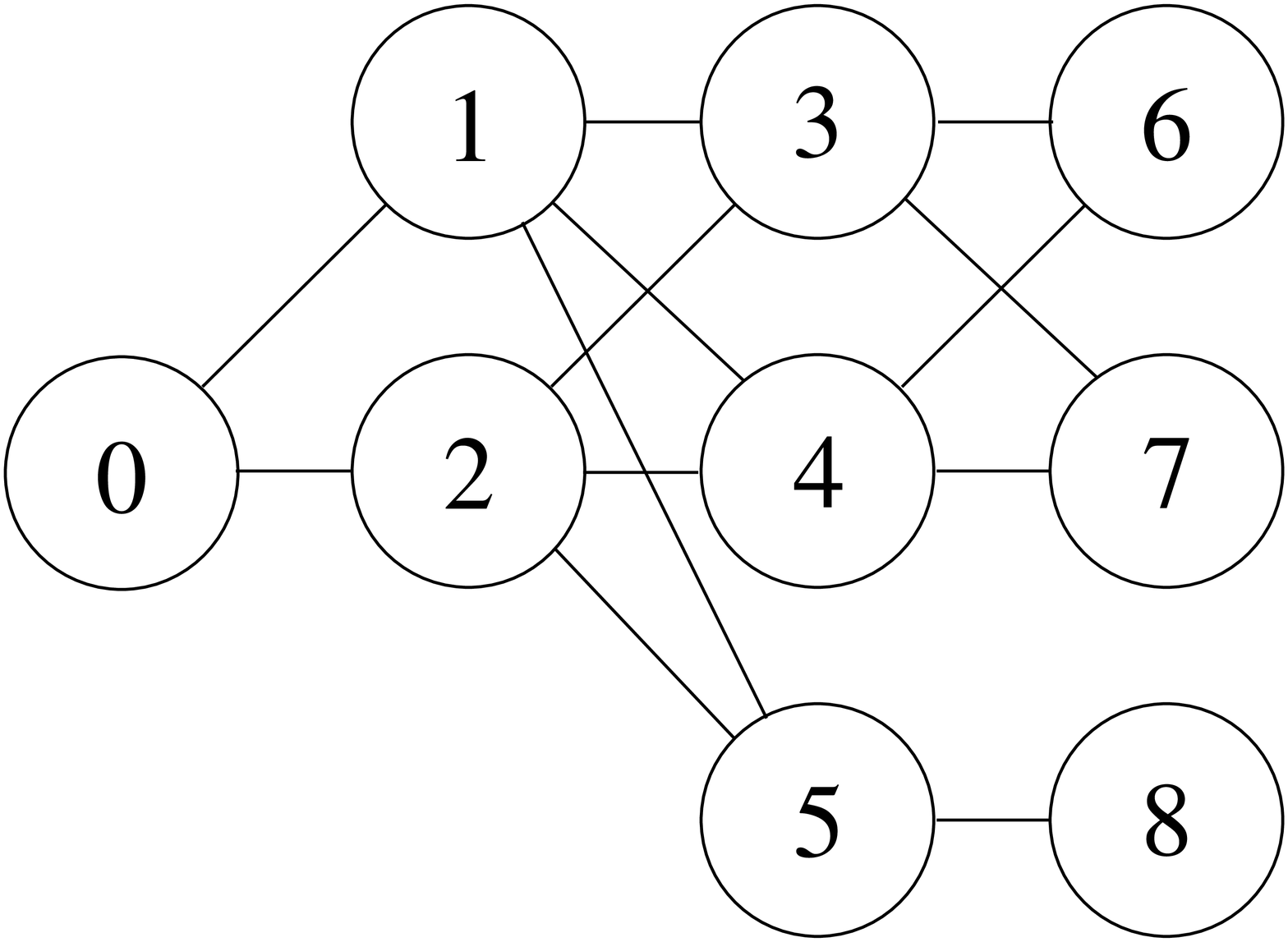}}
\end{center}
\caption{The trajectory graph representing the set of histories in our example.}
\label{fig1}
\end{figure}

Trajectory graphs \cite{Griffiths1993,ChisolmSudarshanJordan1996} can
illustrate the difference between minimal and weak decoherence. The
trajectory graph for our example of minimal decoherence is shown in
Fig. \ref{fig1}. There is a column for each time. The different points
in the column represent the different states the system can be in at
that time in its history. There is a path across the graph for each
history $j\triangleright k\triangleright f$. These paths are also
called trajectories.

\begin{figure}
\begin{center}
\scalebox{0.2}{\includegraphics{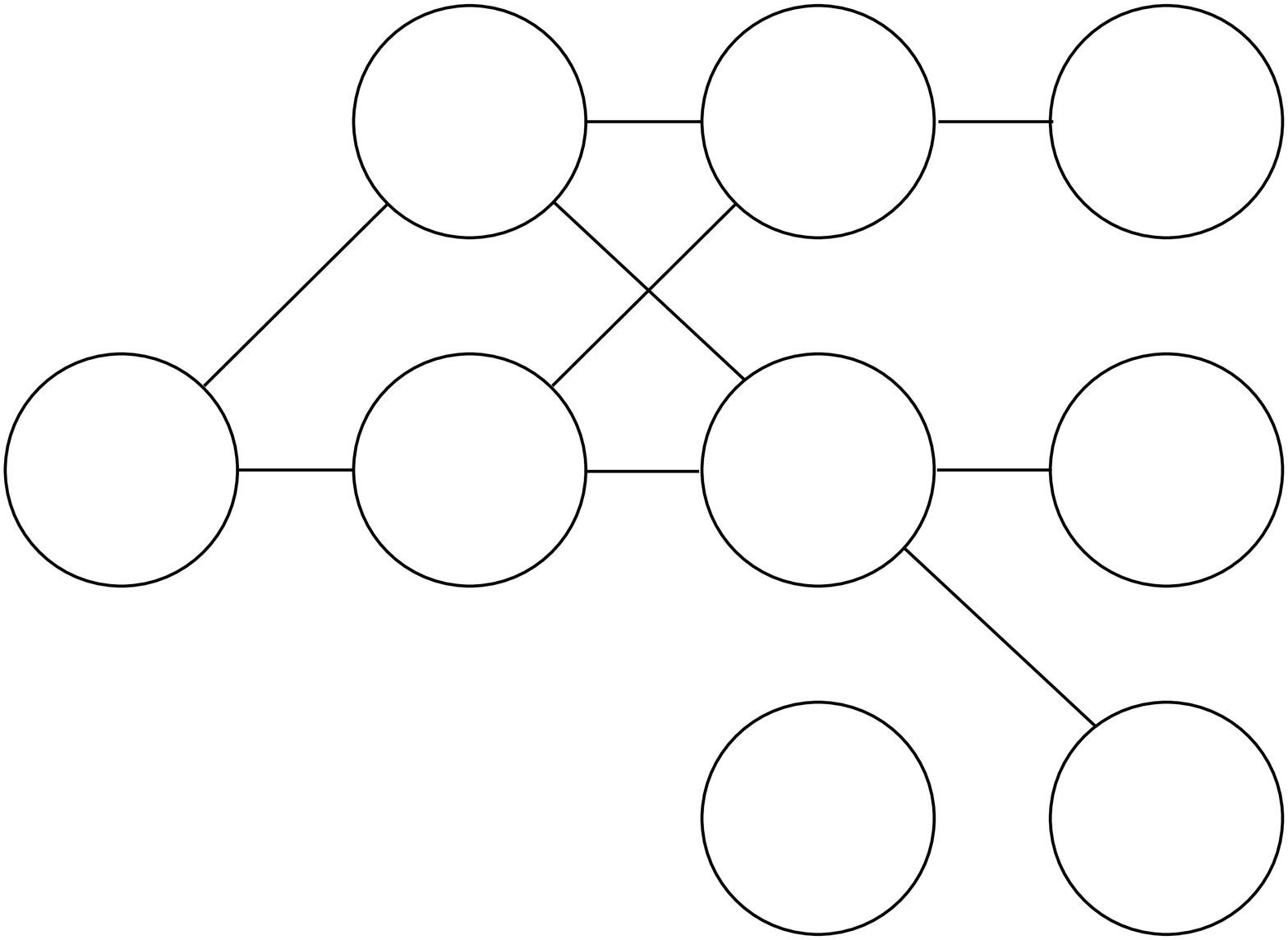}}
\hspace*{0.5in}
\scalebox{0.2}{\includegraphics{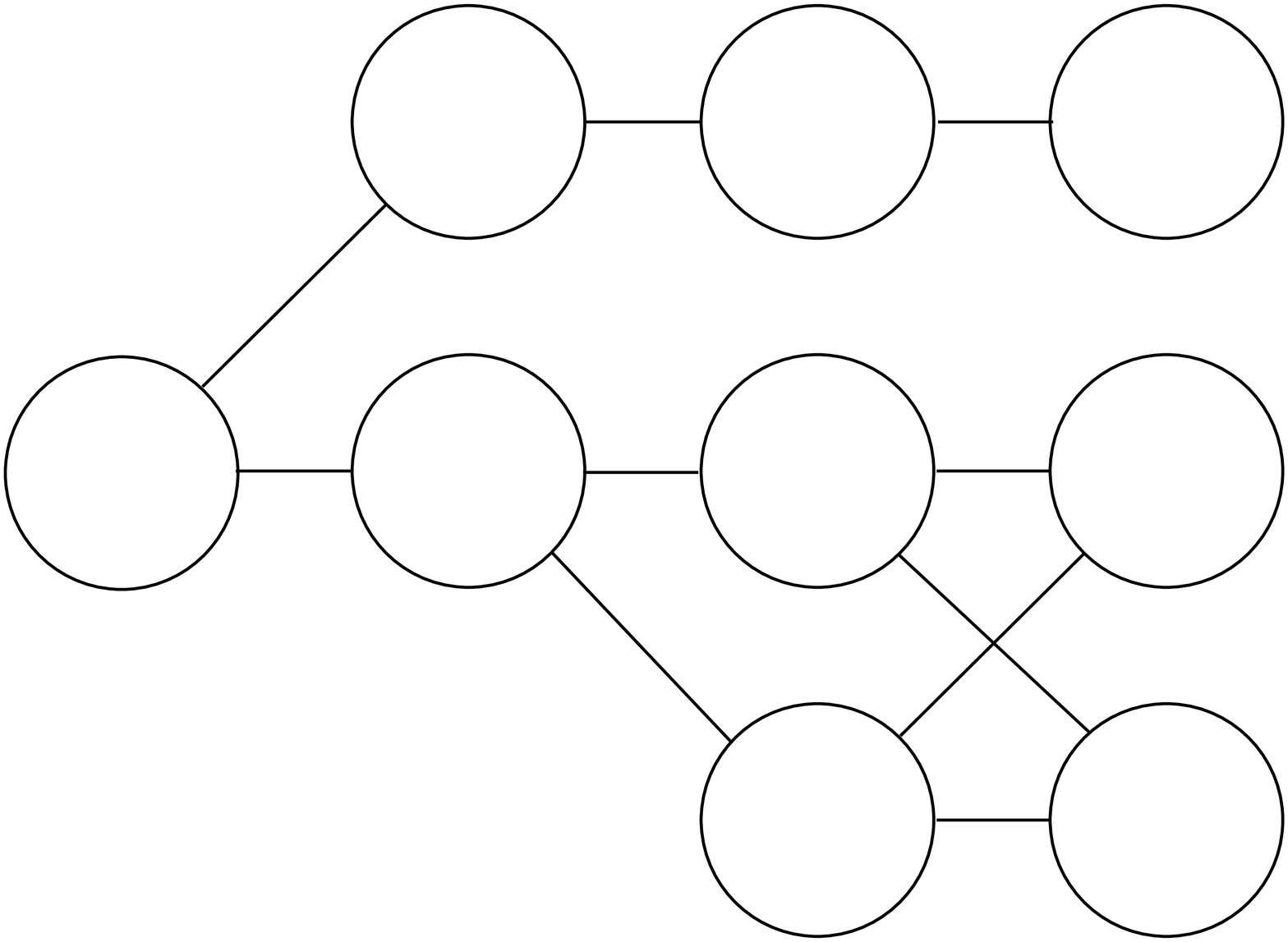}}
\end{center}
\caption{Trajectory graphs allowed by weak decoherence.}
\label{fig2}
\end{figure}

Weak decoherence allows two states at different times to be connected
by at most two different histories
\cite{ChisolmSudarshanJordan1996}. For the numbers of states in our
example, it allows at most two different histories $j\triangleright
k\triangleright f$ and $j'\triangleright k'\triangleright f$ between
$|0\rangle $ and a particular $|f\rangle $. This is because the
condition (\ref{weak}) for weak decoherence requires that the phases
of $\langle 0|j\rangle \langle j|k\rangle \langle k|f\rangle $ and
$\langle 0|j'\rangle \langle j'|k'\rangle \langle k'|f\rangle $ differ
by $\pi /2$, and there can be at most two different complex numbers
with phases that differ by $\pi /2$. Trajectory graphs allowed by weak
decoherence for the numbers of states in our example are shown in
Fig. \ref{fig2}.

Here is a consequence of weak decoherence
\cite{ChisolmSudarshanJordan1996} that is also implied by minimal
decoherence.
\vspace{0.6cm}

\noindent\textit{Theorem}: Projection sum rules imply that a state
$|j\rangle $ does not occur at two different times if the histories
between those times go through a set of states that does not contain
$|j\rangle $.

\noindent\textit{Proof}: The probability for that is the sum of the
probabilities for all the histories that go through $|j\rangle $ at
those two times, which becomes
\begin{eqnarray}
\label{probjkj}
\sum_k\langle f|..1..|j\rangle \langle j|..1..|k\rangle \langle k|..1..|j\rangle \langle j|..1..|0\rangle \langle 0|..1..|j\rangle \langle j|..1..|k\rangle \langle k|..1..|j\rangle \langle j|..1..|f\rangle \nonumber \\
= \sum_k\langle f|j\rangle \langle j|k\rangle \langle k|j\rangle \langle j|0\rangle \langle 0|j\rangle \langle j|k\rangle \langle k|j\rangle \langle j|f\rangle \quad \quad \quad \quad \quad \quad
\end{eqnarray}
when projection sum rules replace the sum over probabilities with the
sum over projection operators
\begin{equation}
\label{unit}
\sum_m |m\rangle \langle m| = 1,
\end{equation}
inserted left and right, for each set of states except the two sets
that contain $|j\rangle $ and one set of states $|k\rangle $ in
between that does not contain $|j\rangle $ but contains at least two
states $|k\rangle $ and $|k'\rangle $ on histories that go through
$|j\rangle $ at both times. The result is zero because the projection
sum rules imply that
\begin{equation}
\label{Rejkj}
\textrm{Re}\langle f|j\rangle \langle j|k\rangle \langle k|j\rangle \langle j|0\rangle \langle 0|j\rangle \langle j|k'\rangle \langle k'|j\rangle \langle j|f\rangle  = 0
\end{equation}
when $|k\rangle $ and $|k'\rangle $ are different. This completes the
proof of the theorem.
\vspace{0.6cm}

\noindent In particular, this does not allow cyclic histories where
there is change.

\section{Discussion}\label{six}

Does it matter that there is a difference between minimal and weak
decoherence? Actual decoherence that is produced physically is more
likely to be medium decoherence \cite{Gell-MannHartle1993}, the
condition that
\begin{equation}
\label{medium}
\langle f|k\rangle \langle k|j\rangle \langle j|0\rangle \langle 0|j'\rangle \langle j'|k'\rangle \langle k'|f\rangle  = 0 \; \, \textrm{unless} \; j=j' \; \textrm{and} \; k=k',
\end{equation}
which is more restrictive than either minimal or weak decoherence. It
is more likely that physical processes will reduce magnitudes to zero
than that they will line up phases to satisfy equations involving real
parts. In the important application of consistent histories to the description of a measuring process in fully quantum terms, the histories are shown to satisfy medium decoherence \cite{Griffiths2002M}.

It is for logical reasons that we want to know the minimum decoherence
structure needed for consistent use of probabilities. Indeed,
Omn\`{e}s has said that
\begin{quote}
when it comes to proving beyond any doubt that a
statement is meaningless whatever its context, ... saying for instance
through which arm of an interferometer a photon goes ... one must make
sure that no possibility has been left out, and one must use necessary
and sufficient conditions. \cite[page 346]{Omnes1992}
\end{quote}
Griffiths has expressed different interests, which have not made it a
priority to pin down the necessary and sufficient conditions:
\begin{quote}
The use of a weak consistency condition has the
advantage that it allows a wider class of consistent families in the
quantum formalism. However, greater generality is not always a virtue
in theoretical physics, and it remains to be seen whether there are
``realistic'' physical situations where it is actually helpful to
employ weak rather than strong consistency. \cite[following
Eq.\ (3.16)]{Griffiths1996}
\end{quote}
\begin{quote}
Even weaker conditions may work in certain cases. The
subject has not been exhaustively studied. However, the [stronger,
essentially medium decoherence] conditions are easier to apply in
actual calculations than are any of the weaker conditions, and seem
adequate to cover all situations of physical interest which have been
studied up till now. Consequently, we shall refer to them from now on
as ``the consistency conditions'', while leaving open the possibility
that further study may indicate the need to use a weaker condition
that enlarges the class of consistent families. \cite[page
142]{GriffithsBookCh}
\end{quote}
There has not been established recognition of what the necessary and
sufficient conditions are.

We have three different logical structures: minimal, weak, and medium
decoherence. Each is well defined and consistent within
itself. Minimal decoherence is all that is necessary. Its requirements
are minimal. It imposes the weakest restrictions and includes the most
histories. It includes all histories that satisfy projection sum
rules. The probabilities in projection sum rules are calculated by
using the Born rule and the von Neumann rule, the most clearly
established rules for the use of probability in quantum
mechanics. Their use in minimal decoherence is always to calculate a
probability for the truth at successive times of propositions
represented by projection operators, for example the
probability (\ref{probjk}), (\ref{probj34}), (\ref{prob12k}), or
(\ref{prob1234}). Projection sum rules relate all the probabilities
that can be calculated that way, the probabilities for all possible
outcomes of measurements made to test those propositions, so there is consistency. History sum
rules contain quantities that are taken to be probabilities but can
not be calculated that way, for example the quantities (\ref{prob?})
and (\ref{prob??}). The operational meaning of these quantities is not
clear to us. They can not be determined by measurements made to test a proposition represented by a projection operator at each
time. Weak decoherence requires history sum rules. It allows fewer
histories, and it brings in new variables that require new
measurements and a new way to calculate probabilities. It is more than
is necessary.

Fortunately, most logical arguments involving weak decoherence use
only a part of it that is included also in minimal decoherence, the
conditions for histories that differ at only one time, like
Eqs.\ (\ref{Rej34}) and (\ref{Re12k}). This is all that was originally
considered \cite{Griffiths1984}. Minimal decoherence implies more; for
example, it implies Eq.\ (\ref{Re1234}).

There is an argument for going in the other direction, from weak to
medium decoherence, rather than from weak to minimal: if weak
decoherence is assumed to apply in a universal manner to systems and
composites of separate and independent systems, it implies medium
decoherence \cite{Diosi2004}. We see no reason to make this assumption,
because we see that weak decoherence is not necessary; minimal
decoherence is enough. If we begin with minimal instead of weak
decoherence, there is no argument that gives medium decoherence like
the argument using composite systems does from weak decoherence, no
argument that bypasses the additional assumptions needed for weak
decoherence. Our example illustrates the difference between minimal
decoherence and weak decoherence. We discuss the assumptions needed
for these. We do not consider stronger assumptions that imply a
stronger form of decoherence which may be suitable for various
purposes.

Clarifications we have achieved highlight questions that remain
open. Can quantum histories be defined broadly enough to be the
elements of a sample space with a probability measure in the sense of
the axioms of probability theory stated by Kolmogorov
\cite{Kolmogorov1950}? Kolmogorov requires that any two disjoint
elements of the sample space can be joined by an ``or'' operation. Can
that be done simply by adding operators? For an array of states
numbered as in our example, the operators $|f\rangle \langle
f|3\rangle \langle 3|j\rangle \langle j|$ and $|f\rangle \langle
f|4\rangle \langle 4|j\rangle \langle j|$ correspond to the histories
$j\triangleright 3\triangleright f$ and $j\triangleright
4\triangleright f$ and can be used as descriptions of these
histories. The sum of these operators,
\begin{equation}
\label{j34opsum}
|f\rangle \langle f|3\rangle \langle 3|j\rangle \langle j| + |f\rangle \langle f|4\rangle \langle 4|j\rangle \langle j| = |f\rangle \langle f|\bigg(|3\rangle \langle 3|+|4\rangle \langle 4|\bigg)|j\rangle \langle j|
\end{equation} 
is used in Eq.\ (\ref{probj34}). It describes the history specified by
propositions at successive times represented by the projection operators $|j\rangle
\langle j|$, \mbox{$|3\rangle \langle 3|+|4\rangle \langle 4|$} and
$|f\rangle \langle f|$, for which Eq.\ (\ref{probj34}) is the
probability. We can take this history to be the ``or'' composition of
the histories $j\triangleright 3\triangleright f$ and $j\triangleright
4\triangleright f$. What is the ``or'' composition of the histories
$1\triangleright 3\triangleright f$ and $2\triangleright
4\triangleright f$? Does the operator
\begin{equation}
\label{prob?sum}
|f\rangle \langle f|3\rangle \langle 3|1\rangle \langle 1| + |f\rangle \langle f|4\rangle 
\langle 4|2\rangle \langle 2| = |f\rangle \langle f|\bigg(|3\rangle \langle 3|1\rangle \langle 1|+|4\rangle 
\langle 4|2\rangle \langle 2|\bigg)
\end{equation} 
describe a history? It does not describe a history with a proposition represented by
projection operator at each time. If there is any ``or'' composition of the
histories $1\triangleright 3\triangleright f$ and $2\triangleright
4\triangleright f$, described by this sum of operators or otherwise,
how would measurements affirm it? How could the probability for it be
understood?

For the ``or'' composition of the histories $j\triangleright
3\triangleright f$ and $j\triangleright 4\triangleright f$,
described by the operator sum (\ref{j34opsum}), we have the
probability (\ref{probj34}) for the truth at successive times of the propositions represented by the projection
operators $|j\rangle \langle j|$, \mbox{$|3\rangle \langle
3|+|4\rangle \langle 4|$} and $|f\rangle \langle f|$. This kind of
probability for a sequence of events was first discussed in the
literature by Wigner \cite{Wigner1963} and by Aharanov, Bergmann, and
Lebowitz \cite{AharanovBergmannLebowitz1964}. Sometimes its form is
imagined to come from successive ``collapses of the wave function,''
but it can be understood simply as the probability for a sequence of
measurements to have a certain result, without any reference to wave
function collapse \cite{Ballentine1998,Griffiths2002C}. Its operational meaning is
clear. Could there be a more general probability formula that applies
to the ``or'' composition of the histories $1\triangleright
3\triangleright f$ and $2\triangleright 4\triangleright f$ as well?
What would the measurement procedure for it be? How could experiments
verify that it gives the right answer?

A tensor-product enlargement of the mathematical structure has been
proposed that allows a projection operator for every proposition that
is to have a probability \cite{Isham94,IshamLinden94}. It is hoped
that this may provide a foundation for a quantum theory of gravity. It
has not answered the questions of operational meaning. Measurements
that could test the propositions represented by all the new projection
operators have not been described. We think that for ordinary quantum
mechanics it may be more reasonable to just use minimal
decoherence. We feel less need to find meaning for all the sum rules
of weak decoherence when we consider that minimal decoherence does not
require weak decoherence and that minimal decoherence provides a
framework for consistent use of probabilities for all the possibilities described by a sequence of propositions represented by projection operators at different times, the probabilities for all the possible
results of measurements that test those propositions.

\bibliography{Decoherence}

\newpage

\end{document}